\documentclass{article}
\usepackage{amsmath}
\usepackage{geometry}
\usepackage{graphicx}
\usepackage{authblk}
\usepackage{hyperref}
\usepackage[utf8]{inputenc}
\title{Insights from the Algonauts 2025 Winners}
\author[1,2]{Paul S. Scotti}
\author[1,3]{Mihir Tripathy}
\affil[1]{Medical AI Research Center (\href{https://medarc.ai}{MedARC})}
\affil[2]{\href{https://sophontai.com}{Sophont}}
\affil[3]{CAMRI, Baylor College of Medicine}
\date{August 14, 2025}

\begin{document}

\maketitle

The Algonauts 2025 Challenge just wrapped up a few weeks ago. It's a biennial challenge in computational neuroscience in which teams attempt to build models that predict human brain activity from carefully curated stimuli. Previous editions (2019, 2021, 2023) focused on still images and short videos; the 2025 edition, which concluded last month (late July), pushed the field further by using long, multimodal movies \cite{algonauts_archive}. Teams were tasked with predicting fMRI responses across 1,000 whole-brain parcels across four participants in the dataset who were scanned while watching nearly 80 hours of naturalistic movie stimuli \cite{algonauts_data}. These recordings came from the CNeuroMod project and included 65 hours of training data—about 55 hours of Friends (seasons 1–6) plus four feature films (The Bourne Supremacy, Hidden Figures, Life and The Wolf of Wall Street). The remaining data were used for validation: Season 7 of Friends for in-distribution tests, and the final winners for the Challenge were who could best predict brain activity for six films in their held-out out-of-distribution (OOD) set.

The winners were just announced and the top team reports are now publicly available. As members of the MedARC team which placed 4th in the competition, we want to reflect on the approaches that worked, what they reveal about the current state of brain encoding, and what might come next.

Before getting into the specific top-performing entrants, here's the common trends across them all:

\section*{Overall observations}
\begin{enumerate}
    \item \textbf{Reliance on pre-trained feature extractors.}
No top team trained their own feature extractors. The universal strategy was to use pretrained foundation models to convert stimuli into high-quality feature representations. The core engineering challenge was how to integrate these features and align them with fMRI brain activity.

    \item \textbf{Multimodality was essential.}
    Following from the first point, every top team used pre-trained models spanning vision, audio, and language. Predicting activity in higher-order associative brain regions in particular required the model to process multimodal features \cite{dascoli_tribe}. 

    \item \textbf{Architecture didn't really matter.}
    First and second place used transformers, third place used RNNs, fourth place was simply convolutions and linear layers (no nonlinearity!), and sixth place used seq2seq transformers. Despite these differences the final leaderboard scores were all extremely tight, and actually the winning teams were decided by how they implemented model ensembling.

    \item \textbf{Ensembling decided the winner.}
    All top teams used ensembling of some sort. Averaging model variants (often with sophisticated per-parcel weighting) was the most effective way to gain noticeable performance improvements. TRIBE seemed to use the most sophisticated ensembling strategy which we think determined their first place finish.
\end{enumerate}

\section*{First Place: TRIBE – A TRImodal Brain Encoder (Meta AI; d'Ascoli, Rapin, Benchetrit, Banville, \& King)}
The Meta AI team led by Stéphane d’Ascoli took first place with TRIBE \cite{dascoli_tribe}, a model that employs the general multimodal strategy that all the Algonauts teams used. TRIBE ingests text, audio and video representations extracted from large pretrained models and fuses them with a transformer to predict cortical responses. Unlike some of the other top teams, TRIBE only used unimodal models and did not use intrinsically multimodal models like Qwen2.5-Omni or InternVL3. Hence, it seems that while combining features from all modalities was essential for all teams, using intrinsically multimodal models was not.

\begin{figure}[h!]
    \centering
    \includegraphics[width=0.54\textwidth]{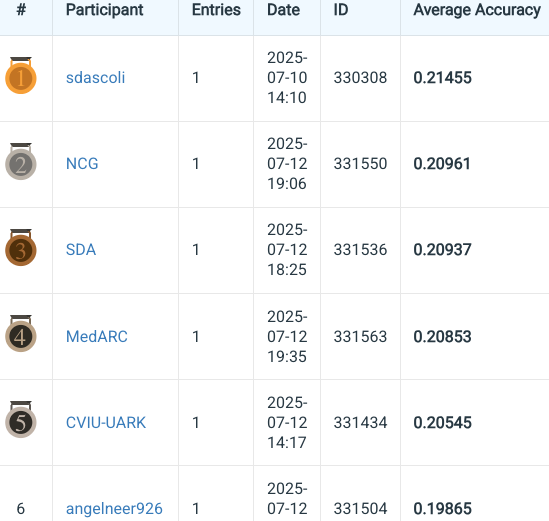}
    \caption{The final Out-of-Distribution (OOD) leaderboard for the Algonauts 2025 challenge.}
    \label{fig:leaderboard}
\end{figure}

Looking at the final leaderboard (Figure~\ref{fig:leaderboard}), TRIBE performed noticeably better than all of the rest in terms of the average Pearson correlation across the 1000 brain parcels. However, digging into the papers a bit more, Eren et al.'s team in 2nd place actually had a submission that scored 0.2125 average accuracy that was not submitted in time. Likewise, our own MedARC paper discusses how we would have placed second with an average accuracy of 0.2117 if we were provided one additional hour to finalize our submissions. In both these cases, the performance increases simply come from better model ensembling. All of this is to say that all the top teams were actually incredibly close in terms of final performance.

In our opinion, the key innovations unique to TRIBE were:

\begin{itemize}
    \item Modality dropout during training, which forced the model to remain robust even when a modality (e.g. audio) was missing. This is a simple and intuitive solution that likely was especially useful given the silent, black-and-white Charlie Chaplin film in the OOD set.
    \item A parcel-specific ensembling scheme: rather than averaging all models equally, they computed validation performance per model per brain parcel and used those scores as softmax weights. The temperature for the softmax can control how aggressively you dial the ranking (e.g., T approaching 0 means winner-take-all, T approaching infinity means pure averaging).
    \item Careful analysis showing that multimodal models systematically outperformed unimodal models in higher-order associative cortices, reinforcing the idea that complex brain regions integrate multiple sensory streams.
\end{itemize}

Further, TRIBE’s report suggested that encoding performance increases with more training sessions (up to 80 hours per subject). However, the trend appears sub-linear and plateauing. In any case, it's not the clean power law seen in large language models and we still look forward to an eventual clean scaling law paper for neuroimaging encoding models.
\begin{figure}[h!]
    \centering
    \includegraphics[width=0.8\textwidth]{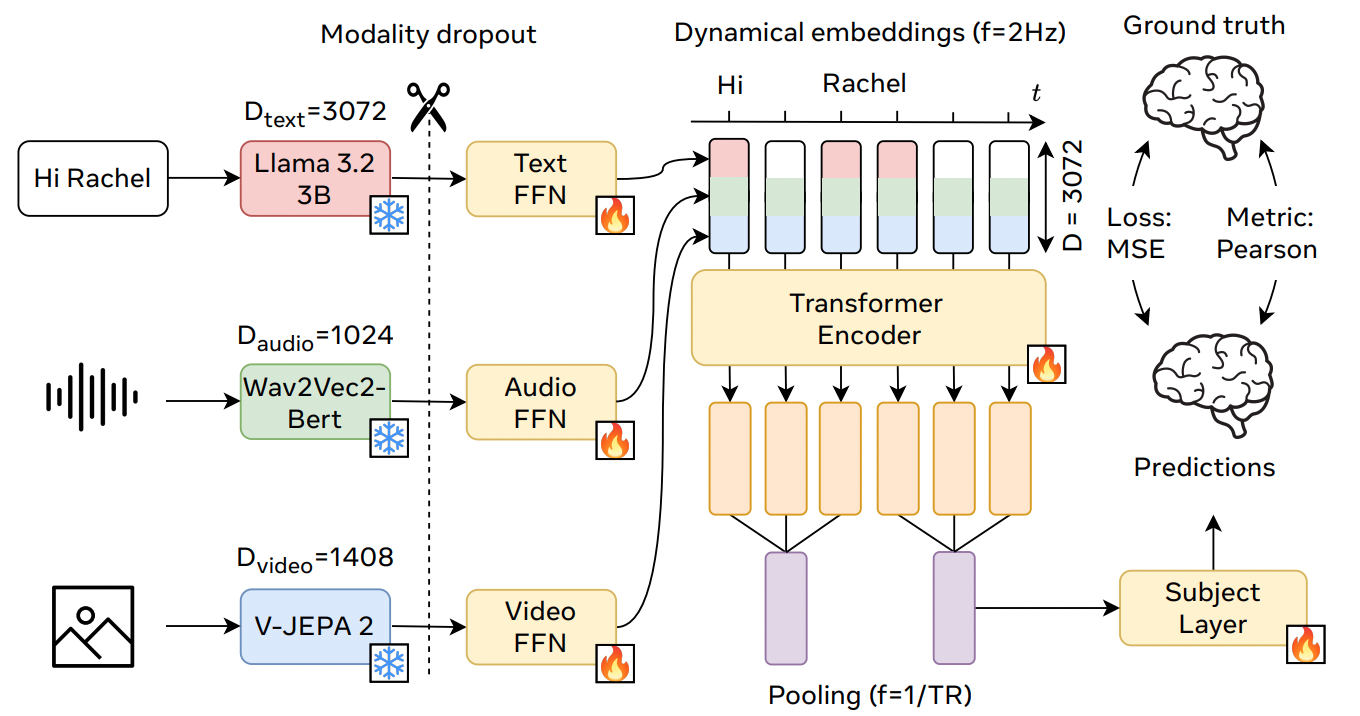}
    \caption{The architecture of Meta AI's TRIBE model, which fused text, audio, and video representations with a transformer to predict cortical responses.}
    \label{fig:tribe_arch}
\end{figure}

\section*{Second Place: VIBE – Video-Input Brain Encoder (Max Planck; Schad, Dixit, Keck, Studenyak, Shpilevoi, \& Bicanski)} 

The NCG team (Max Planck Institute for Human Cognitive and Brain Sciences) placed second with VIBE. VIBE's architecture consists of a modality fusion transformer and a prediction transformer \cite{schad_vibe} which help to separate the challenges of multi-modal feature integration from modeling temporal dynamics of the fMRI time-series. 

Their “modality fusion transformer” integrated features from numerous models (Qwen2.5 and LaBSE for text, BEATs and Whisper V3 for audio, SlowFast, V-JEPA2, \& Qwen2.5-Omni for vision). These features are “fused” via cross attention to create a single, unified representation per time point. Then, their “prediction transformer” models temporal dependencies across time points. Notably, the authors trained the model without a causal mask, allowing it to attend to future time points. This yielded a slight performance increase, suggesting that information about future stimuli helps predict present brain responses.

They maximized performance by ensembling twenty models (which happened to also be the sweet spot for our MedARC submission), achieving a final mean correlation of 0.2125 on the OOD movies. They used separately trained models to predict brain parcels based on the different functional networks they belonged to (e.g., Default Mode Network, Visual).

The authors also experimented with explicitly modeling the hemodynamic response function (HRF). They tested two methods: convolving final predictions with a canonical HRF and using a learnable 1D-convolutional layer. Both approaches decreased performance, leading them to conclude that the model's internal transformer architecture was more effective at learning the nuanced temporal delays on its own than a rigid, predefined HRF. This corresponds to our findings with our models as well.
\begin{figure}[h!]
    \centering
    \includegraphics[width=0.8\textwidth]{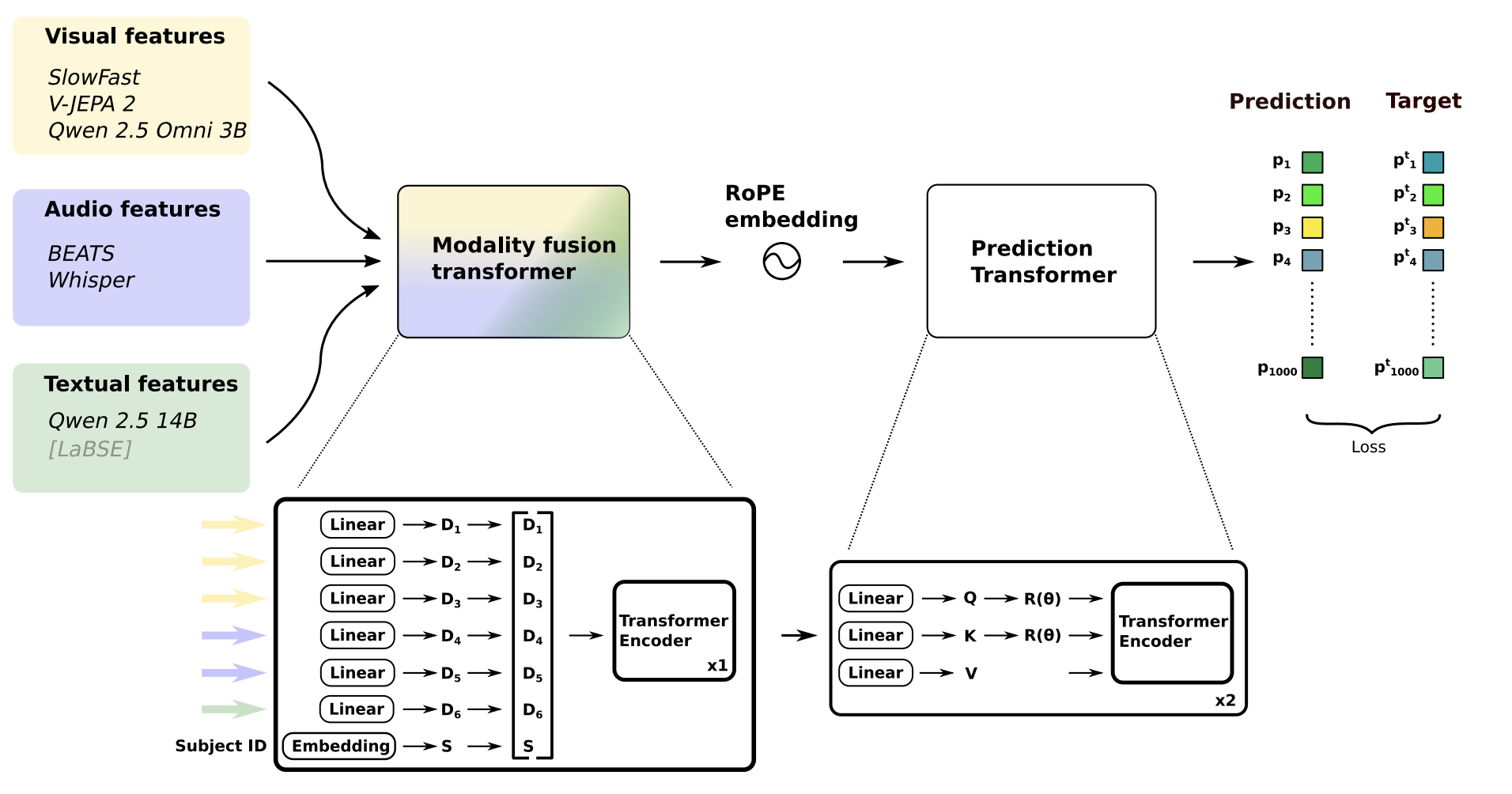}
    \caption{Team NCG's transformer architecture for VIBE.}
    \label{fig:NCG_arch}
\end{figure}

\section*{Third Place: Multimodal Recurrent Ensembles (Max Planck; Eren, Kucukahmetler, \& Scherf)}
The SDA team took a different route with a hierarchical recurrent architecture \cite{eren_ensembles}. They used SlowFast, VideoMAE, Swin Transformer, and CLIP to extract features for visual input, HuBERT and WavLM for speech/audio, CLAP for semantic audio embeddings, BERT for local text semantic features, and Longformer to extract longer-range textual context by prepending the previous episode’s transcripts (if available) when extracting language features.

A key design choice, marking a distinct departure from our own strategy and that of Meta and NCG teams, was how the authors handled temporal information. While we relied on the feature extraction models themselves to capture temporal context by feeding them longer, 20-second clips, the SDA team instead extracted features on a strict TR-by-TR basis. This created a sequence where each time step corresponded to an isolated 1.49-second moment of the stimulus.

These independent feature sequences (visual, audio, and text) were fed into their own dedicated bidirectional Long Short-Term Memory blocks (LSTMs), allowing the use of future and past stimulus features to better understand the context of the current time step when predicting the corresponding fMRI signal. Modality-specific LSTMs modeled temporal dynamics and context within each modality independently. The hidden states from each modality were then fused using simple average pooling. The authors note this straightforward approach was both effective and acted as a regularizer; surprisingly, more complex methods like learned weights or attention did not improve performance.

This fused representation was passed into a second, standard recurrent layer (either an LSTM or GRU) to capture cross-modal dynamics. Finally, lightweight, subject-specific linear heads mapped these features to predict fMRI activity.

Their training strategy included using a brain-inspired curriculum, first optimizing for early sensory regions before gradually shifting to more complex, higher-order association areas as training progressed. This along with a final submission of a 100-model ensemble, built with impressive diversity by training models with different recurrent units (LSTM vs. GRU) and different loss functions, earned them third place in the competition with a final score of $r=0.2094$. Notably, their model achieved the single highest peak-parcel score among all participants (mean $r=0.63$) and performed best on the most challenging subject. We appreciated their unique and effective application of a pure recurrent architecture.



\begin{figure}[h!]
    \centering
    \includegraphics[width=0.8\textwidth]{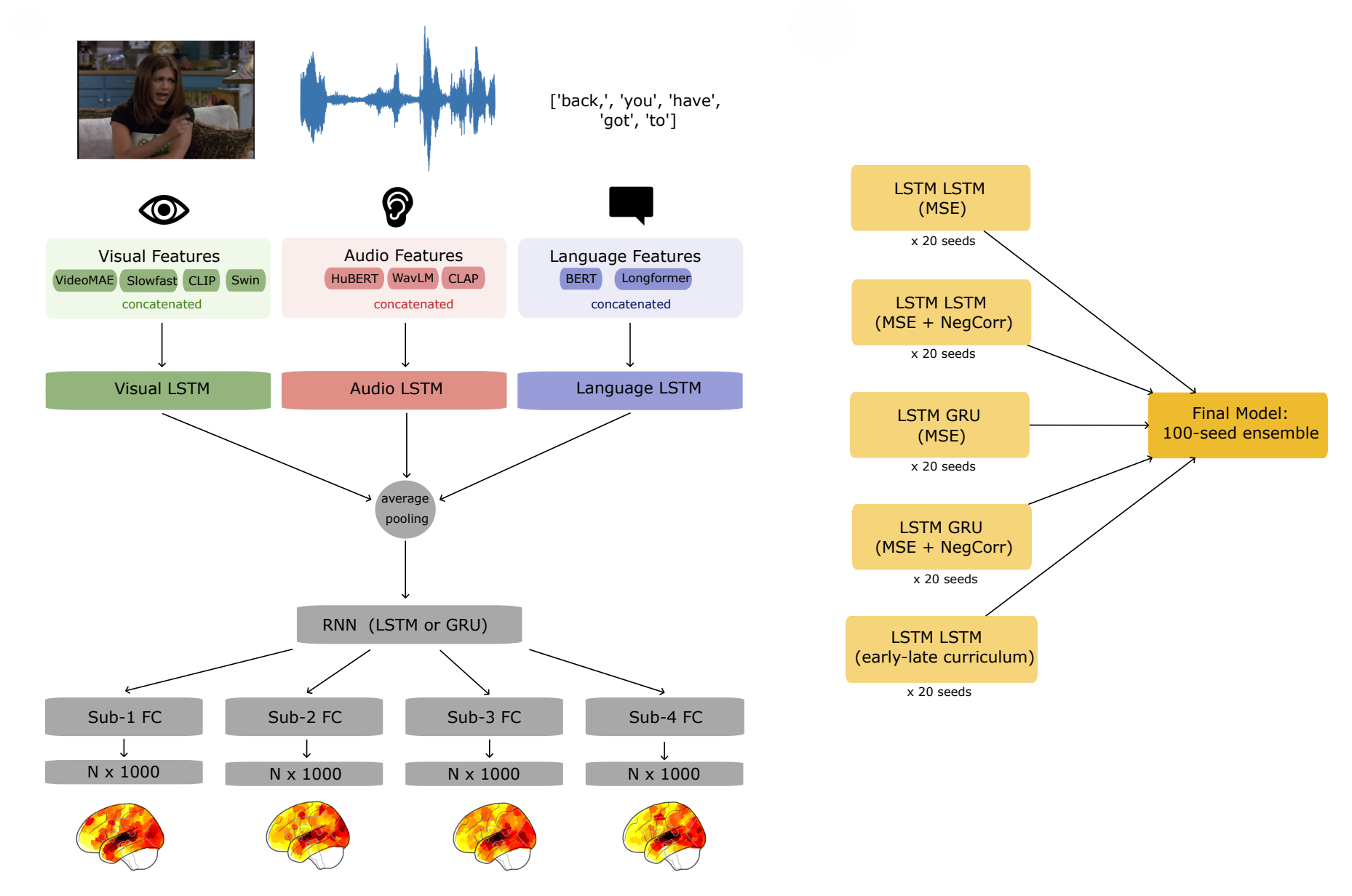}
    \caption{Team SDA's RNN-based architecture \& ensemble strategy.}
    \label{fig:SDA}
\end{figure}

\section*{Fourth Place: Multimodal LLMs and a Lightweight Encoder (MedARC; Villanueva, Tu, Tripathy, Lane, Iyer, \& Scotti)}
Our own MedARC team boasts the most architecturally simple approach. We gathered features from five distinct models—V-JEPA2 (vision), Whisper (speech), Llama 3.2 (language), InternVL3 (vision–language), and Qwen2.5-Omni (vision–language–audio)—and aligned them temporally to the fMRI signals. 

These features fed into a lightweight linear encoder comprising a shared group head plus subject-specific residual heads. The core of our model was 1D temporal convolution, which efficiently captured local time-based patterns, followed by a linear projection. To maximize performance, we trained hundreds of variants and constructed parcel-specific ensembles to increase leaderboard performance. 

The fact that our approach, which was devoid of transformers or any non-linear operations, was so competitive shows that architectural complexity isn't that necessary for modeling neuroimaging data. A simple and efficient linear model can be incredibly effective, suggesting there is significant room for future modeling innovations.

\begin{figure}[h!]
    \centering
    \includegraphics[width=0.8\textwidth]{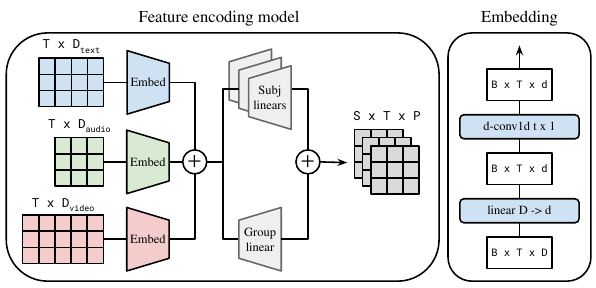}
    \caption{The MedARC team's architecture simply used convolutional and linear layers.}
    \label{fig:medarc_arch}
\end{figure}

\section*{Fifth Place: CVIU-UARK}
The fifth-place team from the Computer Vision and Image Understanding Laboratory at the University of Arkansas (CVIU-UARK) did not release a report.

\section*{Sixth Place: Multimodal Seq2Seq Transformer (Univ. of Chicago; He \& Leong)}
The University of Chicago team introduced a sequence-to-sequence transformer that treated brain encoding as a translation problem \cite{he_seq2seq}. Their model extracted features using VideoMAE, HuBERT, Qwen and BridgeTower, then autoregressively predicted fMRI sequences, attending to both the current stimuli and the history of prior brain states.

A key innovation was their training objective: instead of direct regression, they used contrastive learning, training the model to distinguish the correct fMRI sequence from a set of plausible "distractor" sequences. This forced their model to learn more robust and meaningful neural representations. Further, a shared encoder with partially subject-specific decoders captured common patterns across subjects while allowing individual variation. Although this conceptually novel approach finished just outside the top five, it showed particularly strong performance in higher-order brain regions like the superior temporal sulcus (STS) and default mode network (DMN), demonstrating that autoregressive sequence modeling can handle long-range temporal dependencies in both stimuli and neural responses.

\begin{figure}[h!]
    \centering
    \includegraphics[width=0.8\textwidth]{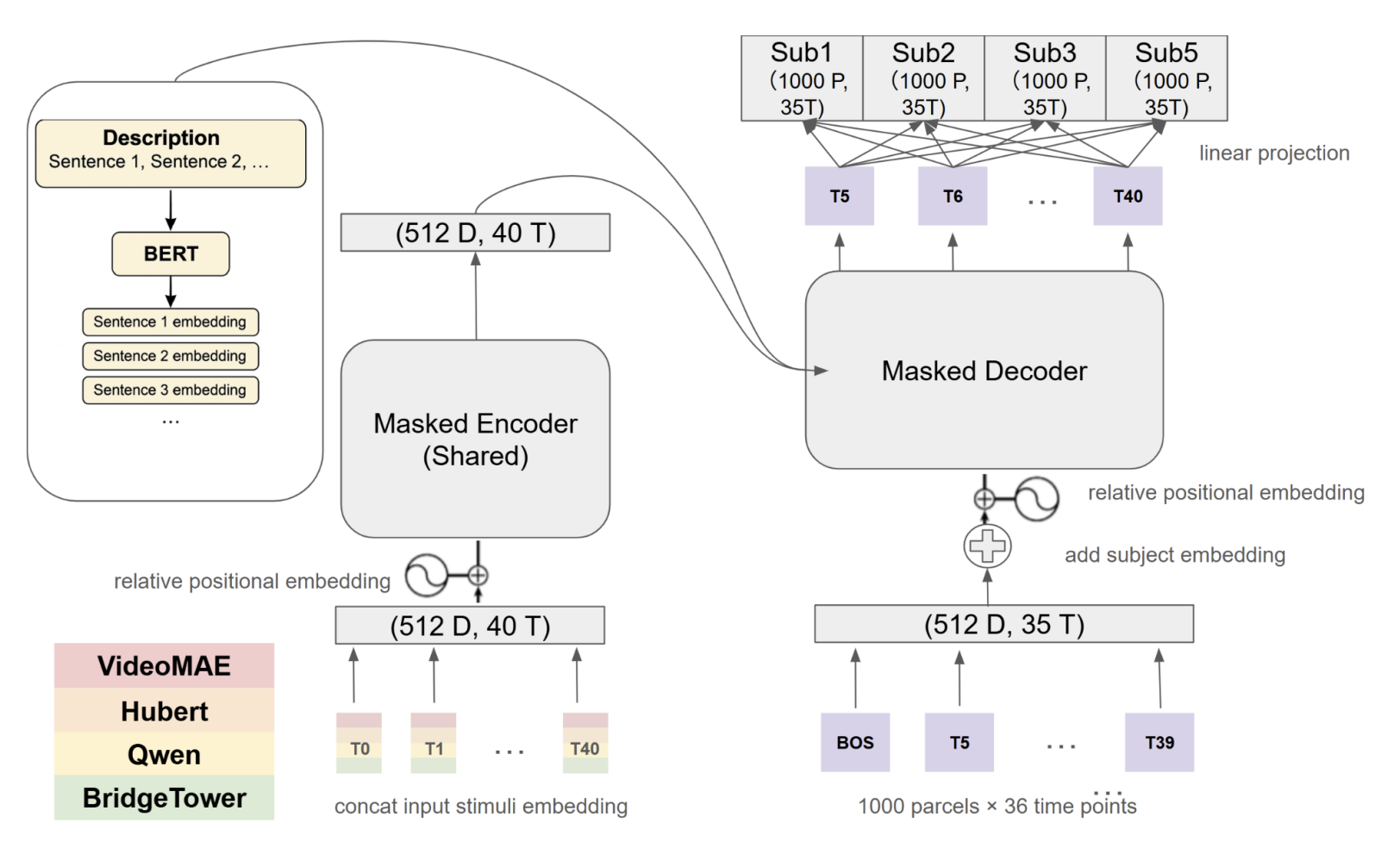}
    \caption{He \& Leong's Seq2Seq architecture.}
    \label{fig:uchicago_arch}
\end{figure}

\section*{Reflections and Looking Forward}
The 2025 Algonauts Challenge delivered impressive models that predict brain responses to rich, multimodal stimuli with unprecedented accuracy. The massive 80-hour dataset of participants watching multiple movies, with careful OOD evaluation, raised the bar for exploring the current best approaches for generalizable encoding models. At the same time, the competition exposed a certain maturity—or stagnation—in the field: all top entries followed a similar recipe of multimodal features, architectural choices did not seem to matter much, and the winner was decided more so by ensembling strategies.

As participants, we find this both encouraging and challenging. Encouraging because we now have various robust pipelines for predicting brain activity in natural multimodal settings. Challenging because breakthroughs may require departing from this pipeline.

For now, the Algonauts Project continues to be a beacon for collaborative, open science. It provides the community with ever richer datasets, rigorous benchmarks and a spirit of friendly competition. We're proud of what we achieved and eager to see how the next generation of Algonauts will continue to push the frontier for computational neuroscience.

\section*{Acknowledgements}
We thank Connor Lane for his helpful comments and edits in drafting this perspective piece.

\end{document}